\documentclass[12pt]{iopart}
\usepackage{epsfig}
\newcommand{\be}{\begin{equation}}
\newcommand{\ee}{\end{equation}}
\newcommand{\bea}{\begin{eqnarray}}
\newcommand{\eea}{\end{eqnarray}}
\newcommand{\bt}{\begin{tabular}}
\newcommand{\et}{\end{tabular}}

\newcommand{\pp}{~~~.}
\newcommand{\vv}{~~~,}

\newcommand{\mnue}{\nu_e}

\newcommand{\rrrc}{\;\;\; \raisebox{1.1ex}{$\lfloor$} \!\!\! \rightarrow}

\newcommand{\ApJ}{{\it Astrophys. J.\,}}

\begin{document}

\title[Supernova Relic Neutrinos in Liquid Argon detectors]{Supernova Relic Neutrinos in Liquid Argon detectors}

\author{A.G. Cocco, A. Ereditato, G. Fiorillo, G.
Mangano \\ and  V.
Pettorino\footnote[3]{alfredo.cocco@na.infn.it\\antonio.ereditato@na.infn.it
\\giuliana.fiorillo@na.infn.it\\
gianpiero.mangano@na.infn.it\\valeria.pettorino@na.infn.it}}

\address{Dipartimento di Scienze Fisiche, Universit\`{a} di
Napoli "Federico II", \\and INFN, Sezione di Napoli, \\Complesso
Universitario di Monte Sant'Angelo, Via Cintia, I-80126 Napoli,
Italy}

\begin{abstract}
We study the possibility of detecting the background of Supernova
Relic Neutrinos (SRN) with liquid Argon Time Projection Chamber
(TPC) detectors. As far as this study is concerned, these
experimental devices are mainly sensitive to electron neutrino
signals, and could provide further information on both Supernova
explosion mechanism and star formation rate at redshifts $z
\leq 1$. We study in details the main contributions to background
in the relevant energy range from $^8$B and $hep$ solar neutrinos
as well as from low energy atmospheric neutrino fluxes. Depending
on the theoretical prediction for the SRN flux we find that for a
3 kton and a 100 kton liquid Argon TPC detectors the signal may be
observed at the 1$\sigma$ and 4$\sigma$ level, respectively, with
5 years of data taking.
\end{abstract}


\maketitle
\eqnobysec


\section{Introduction}

Type II Supernova explosions (SNII) are among the most energetic
events in the Universe known so far. Massive stars with mass
larger than about eight solar masses undergo several burning
phases until the compact core has been fused into iron. Lacking
any further source of nuclear energy to balance gravity, the star
collapses. When the inner core reaches nuclear densities an
outward propagating shock wave is produced, which eventually leads
to a huge luminosity release of the order of $10^{53}$ erg and to
the explosion of the outer layers of the star. This huge energy
flux is almost entirely in the form of neutrinos of all flavors
produced via weak processes which diffuse through the very dense
inner part of the structure and eventually freely escape from the
neutrinosphere, cooling the remnant proto-neutron star on a time
scale of 10 seconds.

This SNII model has been confirmed by the detection of neutrinos from the
SN1987A \cite{1987a1,1987a2}, though the small number of observed
events still leaves important issues unsolved, as for example the shock
revival mechanism, the way the flux is distributed among the neutrino and
antineutrino flavors, and finally a detailed knowledge of the neutrino mean
energies at the neutrinosphere. For a recent review on this and related
issues see {\it e.g.} \cite{janka}.

Future nearby SNII events would probably provide enough
information to achieve a deeper understanding of the core-collapse
mechanism, since many running experiments, such as
Super-Kamiokande \cite{superk}, SNO \cite{sno}, KamLAND
\cite{kamland} and LVD \cite{lvd}, as well as others under
construction, like ICARUS \cite{icarus,vissani,botella3}, will
collect a huge number of neutrino events, of the order of $10^4$.
Waiting for such an event, whose rate is typically estimated to be
of a few per century in our galaxy, it is interesting to study the
possibility to detect with present and future experiments the
isotropic neutrino flux due to all SNII that have occurred so far
in the Universe. This would represent an independent probe of SNII
mechanism, as well as of the SNII rate as a function of redshifts
up to $z \sim 2$. The latter, being simply proportional to the
star formation rate for stars heavier than 8M$_\odot$, is an
important cosmological observable that is presently studied via
optical and UV surveys but that is still poorly known at high
redshifts.

The expected Supernova Relic Neutrino (SRN) flux has been
considered by several authors \cite{srn1,srn2,srn3,srn4,srn5,srn6,srn7}, and the
results span quite a wide range, mainly because of different
modelling of the star formation rate versus $z$. All these
predictions but the one of the simplest {constant supernova rate}
model \cite{srn1} are compatible with the present experimental
upper bound on $\bar{\nu}_e$ flux obtained by the Super-Kamiokande
Collaboration, $\Phi(\bar{\nu}_e) < 1.2$ cm$^{-2}$s$^{-1}$ for
neutrino energies higher than 19.3 MeV \cite{skbound}. A forecast
for future possible detection of the signal in this experiment, as
well as in KamLAND, has been considered in \cite{strigari}. The
authors find that, by adopting a model which is still compatible
with the Super-Kamiokande bound and which is also motivated by UV
density studies and by the Sloan Digital Sky Survey cosmic optical
spectrum bounds on the $z$ behavior of the star formation rate
\cite{bg,glaze}, the SRN background may be detected at the
1$\sigma$ level in a few years running.
Furthermore, it has been recently pointed out that the addition of
a small fraction of gadolinium trichloride in water Cherenkov
detectors would strongly lower the background below 18 MeV
\cite{beacom}, since radiative neutron capture by Gd would allow
antineutrino tagging by the coincidence detection of the reaction
$\bar{\nu}_e +p \rightarrow e^+ +n$. Modifying Super-Kamiokande
detector in this way would lead to a possible detection of SRN at
least at 1$\sigma$ with less than one year of further data (see
\cite{strigari} and References therein).

In this paper we consider the possibility of measuring the SRN
signal by using a different class of experiments based on the
liquid Argon TPC (Time Projection Chamber) technique
\cite{crubbia}. Presently, this experimental method is adopted in
the ICARUS experiment whose final 3 kton configuration (T3000) is
planned to be commissioned at the INFN Gran Sasso Underground
Laboratories in the near future \cite{icarus}. We note that large
mass liquid Argon TPC detectors (100 kton) have been also
considered as powerful experimental devices for next generation
neutrino physics and in particular as SNII neutrino flux detectors
\cite{100k}.

Though with different efficiencies, experiments like ICARUS are
sensitive to all neutrino flavors. As we will discuss in Section~3,
neutrinos and antineutrinos interact via neutral current with Argon
nuclei, as well as by elastic scattering off electrons.  Electron
neutrinos can also undergo charged current interactions, which indeed
are the leading processes with the highest cross section. Electron
antineutrino charged current interactions are typically at least one
order of magnitude smaller. It is worth stressing that the SRN signal
would be mainly detected in this case in its $\nu_e$ component, so
that ICARUS-like experiments provide a complementary piece of
information with respect to Super-Kamiokande or KamLAND, which instead
constrain the $\bar{\nu}_e$ flux. The present bound on low energy
$\nu_e$ steady flux comes from the Mont Blanc Laboratory,
$\Phi_{\nu_e} \leq 6.8 \cdot 10^3$ cm$^{-2}$ s$^{-1}$ in the energy
range 25 $\div$ 50 MeV \cite{montblanc}. As we will see in the
following a liquid Argon TPC experiment like ICARUS might improve this
result by several orders of magnitude.

The main background sources for SRN events, in the relevant
neutrino energy range of 10 $\div$ 50 MeV, are given by solar and
low energy atmospheric neutrinos. These fluxes are indeed large
enough to completely overwhelm the SRN signal at low energies,
lower than the $^8$B neutrino flux endpoint at approximately 16
MeV, as well as for sufficiently high energies where the
atmospheric neutrino flux becomes the dominant contribution.
However, in the intermediate energy range of 16 $\div$ 40 MeV the
expected SRN signal is likely to represent a substantial fraction
of the total flux. By suitably choosing the energy window, fluxes
as the one considered in \cite{strigari} may therefore be
detectable.

This paper is organized as follows. In Section~2 we review the main
features of SRN fluxes and we introduce the fiducial model used in our
analysis. Section~3 covers the neutrino interaction processes in
liquid Argon TPCs. The results of our simulation are reported in
Section~4, where the expected energy spectrum and event rate from SRN
are studied as a function of the star formation rate and the effect of
the resonant neutrino oscillations in the outer Supernova matter
layers. We also discuss there the backgrounds from solar and
atmospheric neutrinos, as well as other possible sources of
background, comparing them with the SRN flux. Finally, we report our
conclusions in Section~5.

\section{Supernova Relic Neutrino flux}

The neutrino differential flux $\Phi_\alpha(E_\nu)$ (number of
neutrinos per energy interval, per unit time and per unit area)
from past SNII exploded in our observable Universe can be written
as follows: \be \Phi_\alpha(E_\nu) = c
\,\int_0^{z_{max}} \frac{dz}{H(z)} R_{SN}(z) \langle
N_\alpha(E_\nu(1+z))\rangle \label{flux} \vv \ee where $\alpha$
denotes the neutrino or antineutrino flavor and $H(z)$ is the
Hubble parameter \be \fl H(z)= H_0 \left(\Omega_m (1+z)^3+
\Omega_\Lambda +\Omega_R (1+z)^4+
(1-\Omega_m-\Omega_\Lambda-\Omega_R)(1+z)^2 \right)^{1/2} \pp
\label{hpar} \ee Since SNII produce equal fluxes for $\mu$ and
$\tau$ neutrinos/antineutrinos, in the following we will
collectively denote these states as $\nu_x$ and $\bar{\nu}_x$. In
Equation (\ref{hpar}) $H_0$ = 100 $h$ km s$^{-1}$ Mpc$^{-1}$ is
the present value of the Hubble parameter, with $h=0.7 {\pm} 0.1$
\cite{hst,wmap}, while $R_{SN}$ in Equation (\ref{flux})
is the SNII rate per unit time and comoving volume and $\langle
N_\alpha \rangle$ is the number of $\nu_\alpha$ emitted per unit
of initial unredshifted energy by a SNII averaged over the stellar
initial mass function and evaluated at $E_\nu(1+z)$, where $E_\nu$
is the energy measured on earth.

In the following we consider a spatially flat $\Lambda CDM$ model
with $\Omega_m=0.3$, $\Omega_\Lambda=0.7$ and $\Omega_R \sim 0$.
The value of $z_{max}$ is typically set by the experimental
threshold on the lowest detectable neutrino energy. In fact, with
increasing $z$ the energy of neutrinos at production, which is
typically of order of 10 MeV, is redshifted towards smaller values
which eventually become difficult to be measured.

The Supernova rate $R_{SN}(z)$ is the star formation rate for
stellar masses larger than 8M$_\odot$. Several models have been
considered in the recent literature \cite{srn1,srn2,srn3,srn4,srn5,srn6,srn7}, whose
predictions for the neutrino flux vary by approximately one order
of magnitude. For our analysis we will consider a fiducial model
as in \cite{strigari} \bea R_{SN}(z) &=& R_0 \,(1+z)^\beta ,
\,\,\,\,\,\,\,\,\,\,\,\,\,\,\,\,\,\, z \leq 1 \nonumber \\
&=& R_0 \, 2^{\beta-\alpha} \,(1+z)^\alpha , \,\,\,\,\, z>1 \vv
\label{ratesn} \eea where the present rate $R_0$ is estimated in
the range $(0.7 \div 4) {\cdot} 10^{-4}$ yr$^{-1}$ Mpc$^{-3}$
\cite{bg}. Limits on the slope parameters have been
found in \cite{glaze} using the Sloan Digital Sky Survey optical
spectrum giving the ranges $2 \leq \beta \leq 3$ and $ 0 \leq \alpha
\leq 2$. In particular, all our results given in the following are
obtained using $R_0 = 2 {\cdot} 10^{-4}$ yr$^{-1}$ Mpc$^{-3}$,
$\beta=2.5$ and $\alpha=1$.  This choice corresponds to an electron
antineutrino flux which saturates the Super-Kamiokande bound
\cite{skbound}.  As suggested by~\cite{strigari}, such SRN flux could
be detected by Super-Kamiokande in the near future.  
In this case, 
the measurement of the electron neutrino event rate by a future LAr experiment,
compared to the predictions discussed in the following,
would provide an important check of the core-collapse
SNII model, in particular of the $\nu_e$ versus $\bar{\nu}_e$ flux
properties.

Corresponding estimates for higher or lower values of $R_0$ can be
obtained straightforwardly, since this parameter linearly enters
the neutrino differential flux. We also discuss in the following
how our findings are affected when both the parameters $\alpha$
and $\beta$ vary in the ranges shown above.

The second important input of Equation (\ref{flux}) is the
neutrino spectrum emitted by a SNII, which should be averaged over
the mass distribution. Actually, all relevant properties which
characterize the neutrino spectrum are only weakly depending on
the mass of the star, at least for not very heavy stars. We will
therefore make the assumption that the mass average can be
approximated by the neutrino spectrum of a $typical$ Supernova.
This spectrum is obtained by numerically solving the explosion
dynamics \cite{janka} and can be parameterized with good accuracy
by a Fermi-Dirac distribution \be N_\alpha(E_\nu) = k_\alpha
\frac{{\cal L}_{SN}}{T_\alpha^4}
\frac{E_\nu^2}{\exp(E_\nu/T_\alpha-\eta_\alpha)+1} \vv
\label{neutrinospectra} \ee where ${\cal L}_{SN}$ is the Supernova
luminosity, $T_\alpha$ the effective $\nu_\alpha$ temperature and
$\eta_\alpha$ is usually known as {\it pinching} parameter.
Finally, the constant $k_\alpha$ is chosen so that the first
momentum of the distribution is normalized to the total energy
emitted in the $\nu_\alpha$ channel. Typical values for these
parameters, which will be adopted in our study, are the following
\bea && T_{\nu_e} =  3.5 \,MeV, \,\,\,\,\,T_{\bar{\nu}_e} = 5\,
MeV,
\,\,\,\,\,T_{\nu_x} =  8\, MeV \vv \nonumber \\
&&\eta_{\nu_e} =  2, \,\,\,\,\,\eta_{\bar{\nu}_e} =  2,
\,\,\,\,\,\eta_{\nu_x} = 1 \pp \label{parsn}\eea We consider
a total luminosity ${\cal L}_{SN}= 3 {\cdot} 10^{53}$ erg that is
equally distributed among the neutrino/antineutrino species.

Neutrino fluxes from a SNII as measured on earth are influenced by
oscillation phenomena. Once produced at neutrinosphere, neutrinos
propagate in the outer star layers and experience two MSW resonances
at different densities (see {\it e.g.} \cite{smirnov}).  The first
resonance governed by atmospheric mass splitting and the value of the
small mixing angle $\theta_{13}$ occurs at $10^3$ g cm$^{-3}$ (10
MeV/E$_\nu)$. At a lower density of $10^2$ g cm$^{-3}$ (10
MeV/E$_\nu$) a second resonance takes place determined by the solar
parameters $\Delta m_{12}^2$ and $\theta_{12}$. For the experimentally
favored Large Mixing Angle (LMA) solution of the solar neutrino
problem this second resonance satisfies the adiabatic condition for
all realistic matter profiles in a SNII. On the other hand, depending
on the neutrino mass hierarchy and the value of $\theta_{13}$, the
higher density resonance occurs in the neutrino or in the antineutrino
sectors and can be fully adiabatic or rather maximally violating
adiabaticity. Correspondingly, the survival probabilities take
different values, as shown in Table~1 where we consider the two
limiting cases of large and small $\theta_{13}$ mixing angle, $\sin^2
\theta_{13} \geq 10^{-3}$ and $\sin^2 \theta_{13} \leq 10^{-6}$,
respectively. For intermediate values of $\theta_{13}$ the survival
probability depends on neutrino/antineutrino energy, but this effect
is completely negligible for the SRN neutrino flux detection we are
interested in, which is only weakly sensitive to the value of
$\theta_{13}$.
\begin{table*}
\begin{center}
\begin{tabular}{|c|c|c|c|c|}
\hline & mass hierarchy & $\theta_{13}$ & $P(\nu_e \rightarrow
\nu_e )$ & $P(\bar{\nu}_e \rightarrow \bar{\nu}_e )$ \\ \hline
I & normal & large & $\sin^2 \theta_{13}$ & $\cos^2\theta_{12}$ \\
II & inverted & large & $\sin^2 \theta_{12}$ & $\sin^2\theta_{13}$ \\
III & normal/inverted & small & $\sin^2 \theta_{12}$ &
$\cos^2\theta_{12}$ \\ \hline
\end{tabular}
\end{center}
\caption{The $\nu_e$ and $\bar{\nu}_e$ survival probability due to
MSW resonances in a SNII, as function of neutrino mass hierarchy and
$\theta_{13}$ mixing angle.} \label{survprob}
\end{table*}

The effect of oscillations on SRN event will be discussed in the following.
We use the LMA solution for solar neutrino problem which gives as best
value $\sin^2 \theta_{12} = 0.3$ \cite{fogli}. Since $\nu_x$ are produced
with a higher mean energy we expect that in case I, which implies that all
detected $\nu_e$ were born at neutrinosphere as $\nu_x$, the number of SRN
events would be larger. In fact, the cross section for charged current
interaction of $\nu_e$ with Argon nuclei, which is the leading interaction
channel, grows with energy more rapidly than neutrino energy to the first
power. The two scenarios II and III give instead very close results.

\section{Neutrino detection with Liquid Argon TPC detectors}

Neutrinos interact in liquid Argon TPCs via charged and
neutral current interactions off Argon nuclei, as well as from
elastic scatterings on atomic electrons, as
described in the following along with Liquid Argon TPC 
detection capabilities.


\subsection{CC interactions}

\begin{figure}[!hbtp]
\begin{center}
\epsfig{file=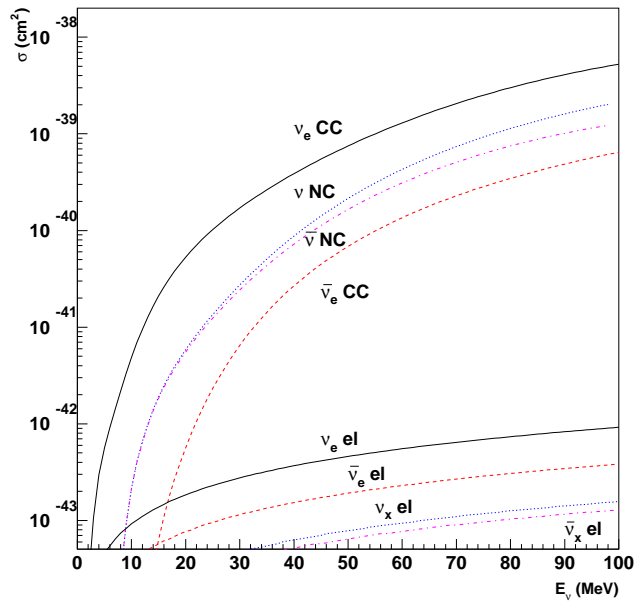,width=9truecm} \caption{The cross section
for neutrino interaction processes in liquid Argon. From top to
bottom, charged current $\nu_e$Ar, neutral current $\nu_e$Ar,
neutral current $\bar{\nu}_e $Ar, charged current $\bar{\nu}_e
$Ar, and elastic scattering for $\nu_e$, $\bar{\nu}_e $, $\nu_x$
and $\bar{\nu}_x$ on electrons. The neutral current results, which
applies to all neutrino flavors, are from \cite{botella3}.}
\label{fig:sigma}
\end{center}
\end{figure}

Electron neutrino charged current interactions on Argon \be \mnue
\ + \ ^{40}{\rm Ar} \ \rightarrow \  ^{40}{\rm K}^*\  + \ e^- \vv
\ee proceed via the creation of an excited state of $^{40}$K and
its subsequent gamma decay. The threshold for this process is
given by the sum of the known $Q$-value of the inverse reaction
(beta decay of $^{40}$K, $Q=1.505$ MeV) and the energy needed to
produce the given excited state of the $^{40}$K. At low energy, as
discussed in \cite{ormand}, the main contributions to the cross
section are from a Fermi transition to the isobaric analog state (IAS) of
$^{40}$K$^*$ and Gamow-Teller transitions to three low-lying
$^{40}$K$^*$ states. Details on the transitions taken into account 
in this analysis are shown in Table~\ref{cctable}.\\
The charged current interaction cross section at higher neutrino
energies (30 MeV $< E_\nu <$ 100 MeV) has been evaluated in
\cite{kolbe} by using a Random Phase Approximation calculation.
Levels up to J=6 have been taken into account. A plot of the cross
section is shown in Figure~\ref{fig:sigma}.

A similar analysis of the $\bar{\nu}_e$ charged current process \be
\bar{\nu}_e \ + \ ^{40}{\rm Ar} \ \rightarrow \ ^{40}{\rm Cl}^*\ + \
e^+ \vv \ee has also been carried out in \cite{kolbe}.  This process
has a higher neutrino energy threshold of 7.48 MeV.  The corresponding
cross section is also reported in Figure~\ref{fig:sigma}. Notice that
charged current cross sections grow with neutrino energy more rapidly
than $E_\nu$ to the first power.
\begin{table}
\begin{center}
\begin{tabular}{|l|c|c|}
\hline
Transition & Excitation level (MeV) & Branching Ratio (\%) \\
\hline
Fermi & 4.384 & 32.76 \\
Gamow-Teller & 3.798 & 13.69 \\
Gamow-Teller & 3.11 & 18.16 \\
Gamow-Teller & 2.73 & 28.94 \\
\hline
\end{tabular}
\end{center}
\caption{\label{cctable} Nuclear excitation levels for $^{40}$K
used in the analysis. See \cite{ormand} for further details.}
\end{table}

\subsection{NC interactions}

For relatively low energy transfers from the neutrino, neutral
current reactions on Argon nuclei proceed via the excitation of
nuclear resonances decaying back to Argon ground state with the
emission of one or more photons \bea \nu_\alpha \ +\  ^{40}{\rm
Ar} \ \rightarrow \   &^{40}{\rm Ar}^*& + \nu_\alpha \nonumber \\
& \rrrc & ^{40}{\rm Ar} + \gamma_1 + \dots \gamma_n \pp \eea A
study of nuclear levels of Argon reveals that the most energetic
photon emitted in this process has an energy not exceeding 11 MeV
\cite{ensdf}. These photons typically produce Compton electrons or
electron/positron pairs, which however have energies lying outside
the energy range we are interested in. In fact, as we will see in
the following, the SRN signal to background ratio is expected to
be maximal by selecting the outgoing electron (positron) energy
range above the threshold given by the steep rising of $^8$B solar
neutrino flux, at approximately 16 MeV. We also note that there is
experimental evidence for a giant quadrupole resonance at 17.7 MeV
\cite{ensdf}, which is likely to produce multi-photon transition
to ground state. In this case too, de-excitation of Argon would
produce electrons with energy smaller than 16 MeV. Finally, for
higher excitation energies, Argon nuclei are no longer stable and
the final state contains one or more nucleons and a remnant
nucleus, which can be distinguished from a single outgoing
electron signal in liquid Argon TPC. As a consequence, in the
following we will not consider neutral current interactions.

\subsection{ES interactions}

In our study we also consider neutrino elastic scattering, whose
cross sections, shown in Figure~\ref{fig:sigma}, have a linear
dependence on neutrino energy $E_\nu$ \bea  \sigma( \nu_e e^-
\rightarrow \nu_e e^- ) &=& 9.20
{\cdot} 10^{-45} (E_\nu/{\rm MeV}) \,\,\, {\rm cm}^2 \vv \\
 \sigma( \bar{\nu}_e e^- \rightarrow \bar{\nu}_e e^- ) &=& 3.83
{\cdot} 10^{-45} (E_\nu/{\rm MeV}) \,\,\, {\rm cm}^2 \vv \\
 \sigma( \nu_{\mu,\tau} e^- \rightarrow \nu_{\mu,\tau} e^- ) &=&
1.57 {\cdot} 10^{-45} (E_\nu/{\rm MeV}) \,\,\, {\rm cm}^2 \vv \\
 \sigma( \bar{\nu}_{\mu,\tau} e^- \rightarrow
\bar{\nu}_{\mu,\tau} e^- ) &=& 1.29 {\cdot} 10^{-45} (E_\nu/{\rm
MeV})\,\,\, {\rm cm}^2 \pp \eea Notice that the above cross
sections are about three orders of magnitude smaller than the one for
$\nu_e$ CC interactions. As we will see in the
following, elastic scattering gives a very small contribution to
the total SRN neutrino event rate in liquid Argon detectors.

\subsection{Event detection}

With the exception of neutral currents, all the processes considered
above include one electron (or positron) in the final state and
typically one or more photons.
The final state is thus mainly electromagnetic.
The excellent properties of liquid Argon as
an electromagnetic calorimeter medium allow to detect
energies down to a few hundreds of keV, therefore no energy threshold
has been considered in our analysis.

In this paper we will refer in particular to the 
ICARUS detector that
combines the features of a bubble chamber as far as the spatial
resolution and the particle identification are concerned, to those
of an electronic TPC. The detector is continuously active and
sensitive, self-triggering and able to perform high-quality
imaging even of low energy events such as those due to the
interaction of neutrinos from stellar collapses~\cite{icarus}.

We will assume in the following that electrons
above 5 MeV can be measured with full efficiency in the detector,
despite the presence of de-excitation photons. The
latter could be possibly used to improve the knowledge of the incoming
neutrino energy.
As reported in~\cite{mudecay}, the electromagnetic energy
resolution in ICARUS TPC can be parametrized as follows \be \frac{\sigma(E_e) }{E_e} =
\frac{11\%}{\sqrt{E_e({\rm MeV})}} + 2.5\% \vv
\label{resol} \ee 
in the neutrino energy range relevant for our analysis.

\section{SRN signal and backgrounds}

\begin{figure}[!hbtp]
\begin{center}
\epsfig{file=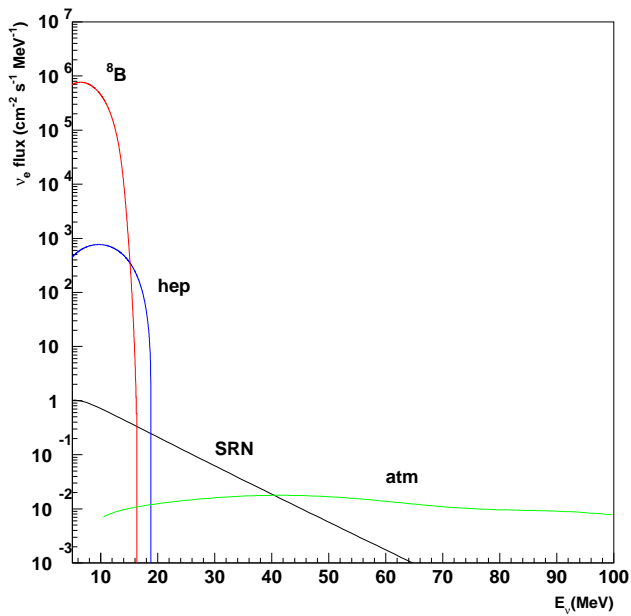,width=9truecm} \caption{The expected SRN
flux of $\nu_e$ for the model of Section~2, a normal neutrino mass
hierarchy and $\sin^2 \theta_{13}=0.02$. The $^8$B and $hep$ solar
fluxes are also shown together with the atmospheric $\nu_e$ flux.
All fluxes are shown versus the neutrino
energy.}\label{fig:theoflux}
\end{center}
\end{figure}

There are two main sources of background for the SRN signal in the
energy range 10 $\div$ 50 MeV. For energies lower than
approximately 20 MeV the (largely) dominant contribution comes
from $^8$B and $hep$ solar neutrinos. These fluxes are quite
accurately known. In Figure~\ref{fig:theoflux} we show the results
of the standard solar model by Bahcall and Pinsonneault
\cite{bahcall}. Total fluxes are chosen according to the best
values suggested in \cite{bahcall}, namely $\Phi_B = 5.79 \,(1
{\pm} 0.23) {\cdot} 10^6$ cm$^{-2}$ s$^{-1}$ and $\Phi_{hep}= 7.88
\,(1{\pm} 0.16) {\cdot} 10^3$ cm$^{-2}$ s$^{-1}$, while the energy
profile has been obtained with the data given in
\cite{bahcalldata} (see also \cite{bahclisi}). The $^8$B flux is
dominating the $\nu_e$ flux up to an energy of 15 MeV, while $hep$
neutrino flux extends well beyond, up to 18.8 MeV.

At higher energies the atmospheric neutrino background becomes
dominant. The estimated flux for $\nu_e$ down to an energy of 10 MeV
at the underground Gran Sasso Laboratory is shown in
Figure~\ref{fig:theoflux} \cite{thank}, as predicted by the FLUKA
Monte Carlo package \cite{fluka}. There is a systematic error in the
evaluation of the atmospheric neutrino flux, due to uncertainties both
in the absolute flux of primary cosmic rays producing neutrinos and in
the cross section for hadronic interactions. We assume in the
following a 30$\%$ relative uncertainty as a conservative estimate for
these systematic effects \cite{thank}.

Assuming the reference model considered in Section~2, we find an
interesting energy window where the SRN signal can be detected, at
least in principle. From Figure~\ref{fig:theoflux} we notice that from
the $hep$ endpoint at 19 MeV up to energies as high as 40 MeV the SRN
$\nu_e$ give a relevant contribution to the total expected
flux. Actually, this prediction strongly depends on the estimated SRN
flux. For those models which predict a lower flux this energy window
can become much smaller or even disappear. In any case, we see that
any attempt to detect the SRN signal, in particular in its $\nu_e$
component, requires a preliminary analysis of the possible signal to
background ratio in the 20 $\div$ 40 MeV energy range.

\noindent Other possible backgrounds have been considered:
\begin{enumerate}
\item 
Electrons coming from the decays of low momentum muons produced in the
interactions of atmospheric $\nu_\mu/\bar{\nu}_\mu$ and escaping
detection before the decay vertex (``invisible muons''), could mimic
neutrino interactions in the target.  This is an irreducible
background in experiments like Super-Kamiokande, where muons with
kinetic energy lower than $50$ MeV are below threshold for emitting Cherenkov photons.
Differently from water Cherenkov detectors, LAr TPCs do not suffer
from this background, being able to detect very low energy tracks,
down to a few hundreds of keV.  Furthermore, in these detectors the
decay of a ${\cal O}$(20 MeV) momentum muon results into an event
topology that can be easily disentangled from a signal event (see for
example~\cite{mudecay}). These events in fact present a high
charge density at the beginning of the track due to the high
ionization of the muon at the end of its range.

\item Beta decays following spallation by cosmic ray muons has also
been investigated as a source of background. Among possible spallation 
products the highest decay energies have been found to be less than 
10 MeV ($^{36}$P $\beta^-$ decay),
well below the lower energy cut used in
the following analysis for SRN searches with LAr detectors.
These processes, thus, do not represent a possible source of background. 
It is worth pointing
out that the very good electromagnetic energy resolution of ICARUS-like
detectors (6\% at 14 MeV \cite{mudecay}) allows for an accurate measurement 
of the visible energy, which is not the case for water Cherenkov detectors
in the same energy range.
Finally we also notice that,
since LAr detectors are fully sensitive to any incoming charged
particle, there is no chance for muon spallation inside the detector
to produce the signature of a single low energy electron.

\item Nuclear recoils arising from the NC atmospheric neutrino
 interactions and from the scattering of fast neutrons coming from the
 surrounding materials could be misidentified due to quenching in LAr.
 We notice that a significant recombination is possible for highly
 charged heavy particles in presence of electric fields of the order
 of 0.5 kV/cm~\cite{quench}; however, the largest Argon nucleus recoil
 kinetic energy for a scattering with a fast neutron is $
 E_{\mbox{\scriptsize recoil}}= 4 E_{\mbox{\scriptsize n}} A/(A+1)^2$
 where A=40.  This means that, assuming a 60\% quenching, a 1 GeV
 neutron is needed to produce a recoil of 50 MeV, 
 quenched to 20 MeV.  At this energy scale for the
 incident particle, inelastic scattering processes take place and the
 outgoing fragments are visible in a LAr detector. In addition, the
 different range out properties of a nuclear recoil and of an electron
 would result in events of very different topologies.
\end{enumerate}
For the above considerations, the only sources of background included
in our analysis are solar and atmospheric neutrino induced events.  

\subsection{MonteCarlo simulation}

We have simulated $10^4$ SRN events according to a SNII spectrum
whose parameters are chosen as in Equation (\ref{parsn}) along with
$10^4$ and $10^5$ solar and atmospheric neutrino events,
respectively. Incoming neutrino spectra have been normalized to the
expected fluxes.  The SRN flux has been normalized according
to~\cite{strigari}. We consider a normal neutrino mass hierarchy and a
large value for the $\theta_{13}$ mixing angle (case I of
Table~1). Interaction of these fluxes with the detector proceeds via
charged current and elastic scattering as described before. We use as
detection signature the electron (positron) produced in the charged
current interaction, or the recoil electron for elastic
scattering. The electron energy $E_e$ spectrum is produced
by using the corresponding differential cross section $d\sigma/dy$,
where $y$ is the fraction of the incoming neutrino energy carried by
the scattered electron ($y=E_e/E_{\nu}$). For CC events we take into
account the thresholds and detection efficiency described in
Section~3. All results are reported in terms of the electron/positron
{\it measured} energy.  For this purpose, the energy resolution of the
detector reported in Equation~\ref{resol} has been taken into
account.  

\begin{figure}[!hbtp]
\begin{center}
\epsfig{file=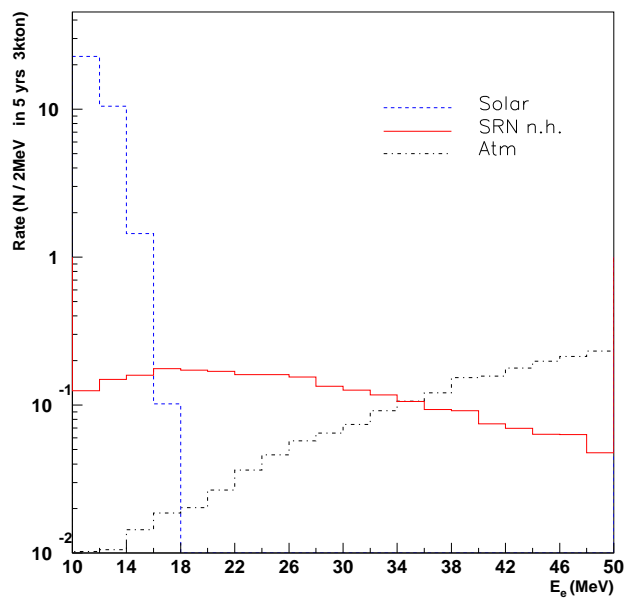,width=9truecm} \caption{The
simulated electron (positron) event spectrum due to solar
neutrinos (blue dashed), atmospheric flux (black dot-dashed) and
SRN neutrinos (red solid), versus the electron (positron) energy.
A normal neutrino mass hierarchy (n.h.) and large $\theta_{13}$ is
assumed. Results are for a 3 kton
detector.}\label{fig:eventvsenergy}
\end{center}
\end{figure}

\begin{figure}[!hbtp]
\begin{center}
\epsfig{file=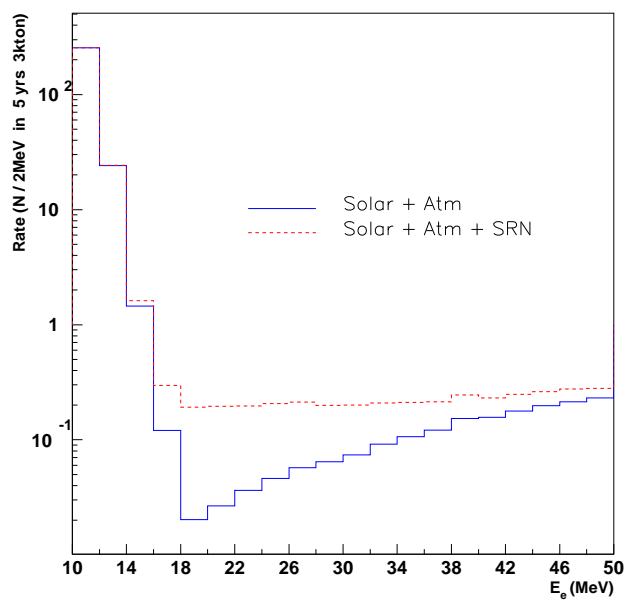,width=9truecm} \caption{The total
event spectrum compared to background due to solar plus
atmospheric neutrinos for a 3 kton detector with a normal neutrino
mass hierarchy and large $\theta_{13}$.}\label{fig:signvsbackg}
\end{center}
\end{figure}

Results are shown in Figure~\ref{fig:eventvsenergy}, where the
number of events due to solar, atmospheric and SRN neutrinos is
given as a function of the electron (positron) energy, in the
energy window 10 $\div$ 50 MeV. The signature of SRN neutrino
induced events for 16 $\leq E_e \leq$ 40 MeV is more clearly seen
in Figure~\ref{fig:signvsbackg} where we compare the total event
energy spectrum with the one expected for the solar and
atmospheric backgrounds only.

\begin{figure}[!hbtp]
\begin{center}
\epsfig{file=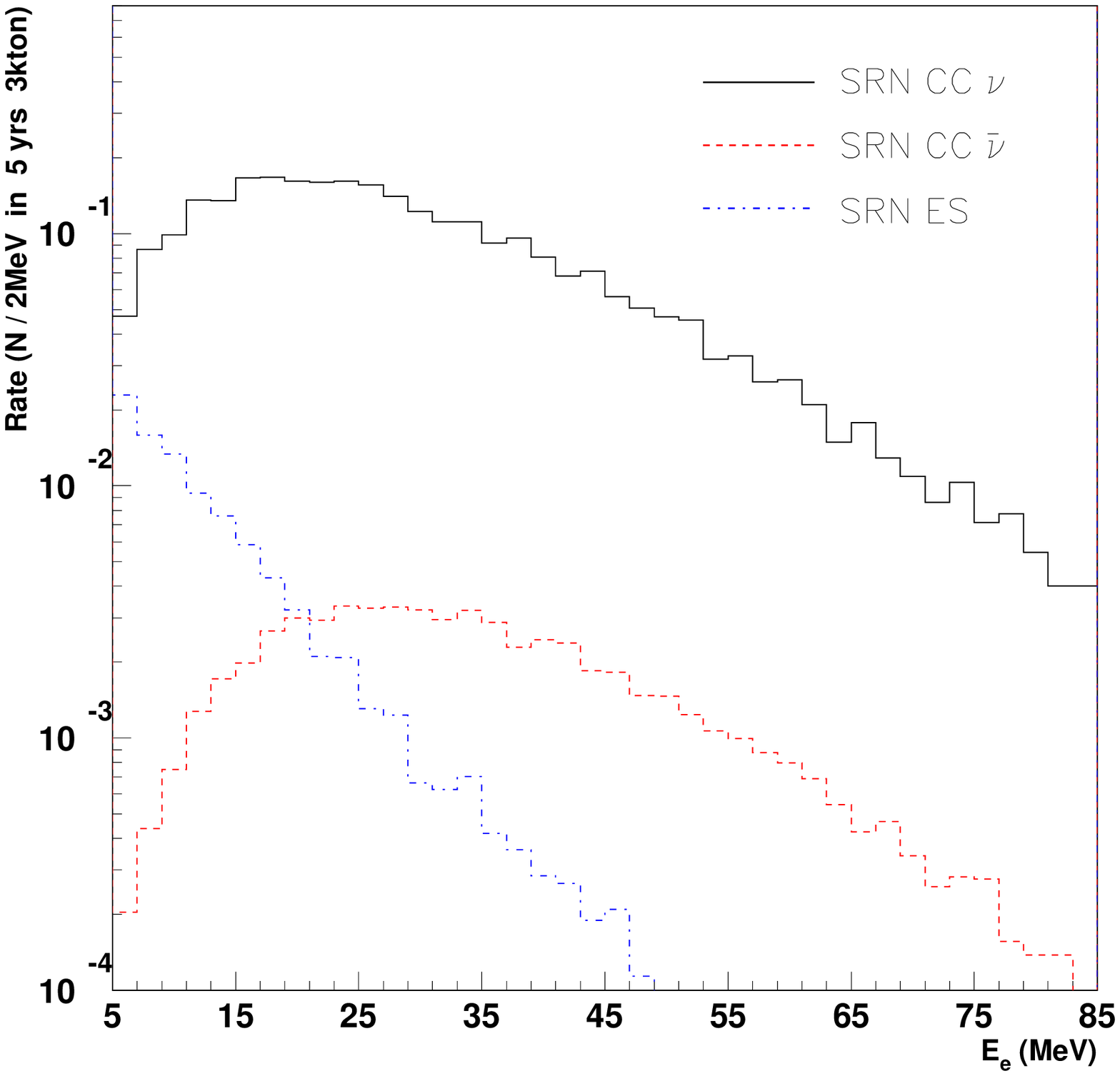,width=9truecm} \caption{The relative
contribution of $\nu_e$ charged current (black solid line),
$\bar{\nu}_e$ charged current (red dashed) and elastic scattering
summed over all neutrino species (blue dot-dashed) for each
electron (positron) energy bin for a 3 kton detector.}
\label{fig:compsigm}
\end{center}
\end{figure}

In Figure~\ref{fig:compsigm} we show the different contributions
to the energy spectrum given by $\nu_e$ and $\bar{\nu}_e$ charged
current interactions and elastic scattering, summed over all
neutrino/antineutrino species. The plot shows that the SRN
signal at an ICARUS-like detector is almost entirely due to
$\nu_e$ charged current interactions.

\begin{figure}[!hbtp]
\begin{center}
\epsfig{file=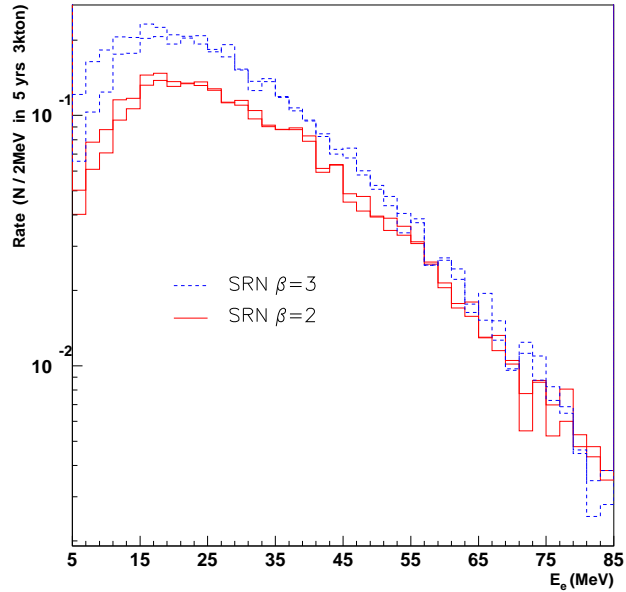,width=9truecm} \caption{The dependence
of the SRN signal on $\alpha$ and $\beta$ as expected in a 3 kton
detector. Upper (blue dashed) curves are for $\beta=3$ and, from
bottom to top, $\alpha=0,2$. The lower (red solid) curves are for
$\beta=2$ and the same values of $\alpha$.}\label{fig:alphabeta}
\end{center}
\end{figure}

A change of the values of $\alpha$ or $\beta$  affects the
expected signal quite differently.  Since the solar neutrino
background is too high for neutrino energies lower than 16 MeV,
for a typical SNII spectrum and even in the best case of a maximal
$\nu_x-\nu_e$ oscillation the contribution of all SNII at redshift
larger than 1 cannot be detected. In fact, for a $\nu_x$ effective
temperature at the neutrinosphere of 8 MeV, we estimate that for
$z=1$ a 15$\%$ fraction of the total $\nu_e$ flux produces
electron charged current events with energy above the 16 MeV
threshold, while this fraction is only 1$\%$ for $z=2$ sources.
This implies that the expected signal is very weakly depending on
the star formation rate at redshift larger than 1. This feature is
clearly seen in Figure~\ref{fig:alphabeta} where we show how the
SRN spectrum versus the electron (positron) energy changes when
varying the star formation slope parameters. The dependence on
$\beta$ is more pronounced. The total number of SRN events changes
by a factor ${\pm} 20\%$ with respect to the result for
$\beta=2.5$, when $\beta$ varies in the range $2 \leq \beta \leq
3$.

\subsection{Results} 

If one integrates the energy spectrum over the selected energy
window for a 3 kton liquid Argon detector like ICARUS running for
5 years one obtains $N_{SRN}=1.7$ events from SRN flux, to be
compared with $N_{BG}= 0.9$ events from the solar and atmospheric
neutrino background. 
To evaluate the error on the expected number of events we consider 
the events in each bin distributed according to Poisson statistics
$\sigma^2_{stat}=N_{SRN}+N_{BG}$ and we add
in quadrature a systematic error $\sigma_{syst}$ to account for a
30$\%$ uncertainty on the atmospheric flux normalization at low
energies and the uncertainty on $hep$ solar neutrino flux as given in
\cite{bahcall}. We finally obtain $N_{SRN} = 1.7 {\pm} 1.6$, for
$16\,{\rm MeV} \leq E_{e} \leq 40 \,{\rm MeV}$

In case of no event detected and applying the approach described 
in~\cite{FC} this would result in an upper limit on the SRN electron 
neutrino flux of
\[ \Phi_{\nu_e} < \ 1.6 \ \ {\rm cm^{-2}\ s^{-1}} \ \ {\rm at} \ 90\% \ {\rm C.L
.} \] This limit would considerably improve the MontBlanc
result~\cite{montblanc}. 
The sensitivity of a LAr experiment like
ICARUS is therefore close to the recently published Super-Kamiokande
bound on the SRN electron antineutrino. 
The statistical significance of a positive observation would be
greatly enhanced by using the full spectral information of the signal
events.  

The above result quite strongly depends on the selected energy
window. If we try to push the lower bound towards lower energies the
steep rise of the $hep$ and $^8$B neutrino events significantly raises
the background. As an example, we find $N_{SRN}=1.8$ and $N_{BG}=2.4$
for 14 MeV $\leq E_{e}\leq $ 40 MeV. Similarly, a larger upper energy
cut gives a less significant result, due to the larger atmospheric
neutrino flux. For 16 MeV $\leq E_{e}\leq $ 50 MeV we find $N_{SRN}=2$
and $N_{BG}=2$ which is slightly better than the result obtained with
a lower energy interval, since atmospheric neutrino flux growth with
energy is not dramatic.

We note that due to the small number of signal events the total
uncertainty is largely dominated by the statistical error, that
contributes for 99$\%$ of the total error. The remaining 1$\%$ is
given by the estimated uncertainty on the atmospheric neutrino
flux. Therefore, we expect that a sensible improvement of the SRN
detection capability of the liquid Argon TPC technique should come
from a larger fiducial mass or a longer running time. Recently,
ideas about next generation liquid Argon TPC detectors have been
put forward \cite{100k}. Fiducial masses as large as 50 $\div$ 100
kton have been envisaged. In this case, for a 100 kton running
for 5 years one would get a more than 4$\sigma$ measurement of the
SRN flux \be N_{SRN} = 57 {\pm} 12, \,\,\,\,\,\,16\,{\rm MeV} \leq
E_{e} \leq 40 \,{\rm MeV} \vv \label{largemass} \ee for a normal
neutrino mass hierarchy and a large $\theta_{13}$ and according to
the fiducial model considered in this paper.

\begin{figure}[!hbtp]
\begin{center}
\epsfig{file=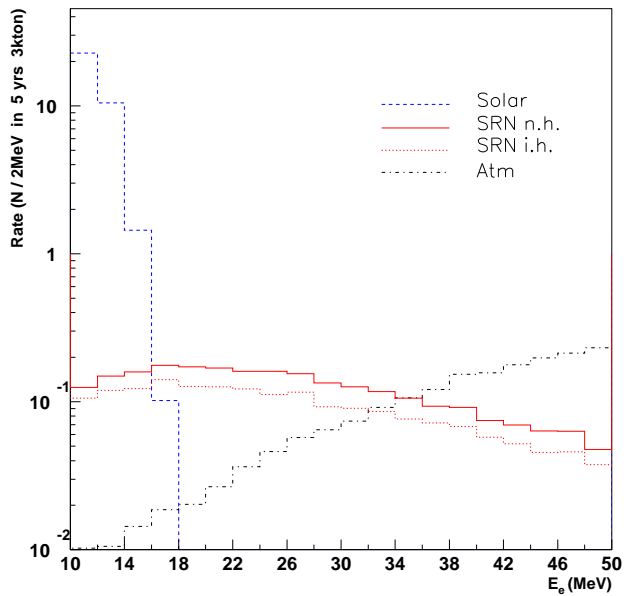,width=9truecm} \caption{The electron
(positron) spectrum for a neutrino normal mass hierarchy and large
$\theta_{13}$ mixing angle (upper red solid curve) and for
inverted mass hierarchy (i.h., lower red dotted curve) and/or
small $\theta_{13}$. The simulation is for a 3 kton
detector.}\label{fig:oscill}
\end{center}
\end{figure}

As a final remark we consider the effect of the different neutrino
oscillation schemes summarized in Table~1. In Figure~\ref{fig:oscill}
we show the expected SRN event spectrum for normal mass hierarchy and
large $\theta_{13}$ and for inverted mass hierarchy and/or small
$\theta_{13}$. In particular the expected event rate in the energy
range 16 $\div $ 40 MeV for cases II and III is now $N_{SRN}=43 {\pm}
12$. As expected, this value is smaller than for scenario I (Equation
(\ref{largemass})), since the electron neutrino flux is shifted toward
lower energies. We see that, provided the star
formation rate is independently fixed by other observations, the
different neutrino mass hierarchy may be distinguished by SRN
observation at the level of 1$\sigma$.

\section{Conclusions}

In this paper we have studied the possibility of detecting the SRN
flux signal at experiments like ICARUS, based on a liquid Argon TPC
technique.  In the relevant energy range, the leading interaction is
the charged current $\nu_e$ scattering on Argon nuclei, which is
revealed via the observation of the produced electron with full
detection efficiency. In contrast with experiments such as
Super-Kamiokande or KamLAND (which essentially detect the interactions
of $\bar{\nu}_e$), the SRN flux would therefore be constrained in the
$\nu_e$ component by this experimental technique, which is thus to be
considered as complementary to the cited ones.

The main detection problem is represented by competing
backgrounds. Indeed, below 16 MeV the solar neutrino flux, in its
$^8$B and $hep$ components, is larger than the expected SRN signal
by orders of magnitude. Similarly, at high energies ($>$40 MeV)
the atmospheric neutrino flux is largely dominating the total
neutrino flux. Nevertheless, we have seen that, mainly depending on
the star formation rate at $z\leq 1$ and more weakly on the
neutrino oscillation pattern inside the SNII, there can be an
electron energy window from 16 MeV to 40 MeV where the SRN
neutrino flux is larger than backgrounds. 

Using a reference model for SRN flux which is presently compatible
with the experimental bound on $\bar{\nu}_e$ flux from relic SNII
obtained by Super-Kamiokande, and which is also well motivated by
astronomical observation, the SRN flux may be observed by a 3 kton
detector like ICARUS at the 1$\sigma$ level with five years of data
taking.  Because of the very few events expected in this case, the
main source of uncertainty is due to the statistical error. A larger
statistical significance will be obtained by next generation, large
mass liquid Argon TPC detectors, reaching the level of 4$\sigma$ for
an exposure of 500 kton $\times$ yr.

\ack

We kindly acknowledge G. Battistoni, A. Ferrari, T. Montaruli and
P. R. Sala, for providing us the low energy atmospheric neutrino
flux, simulated with FLUKA Monte Carlo package. We also thank F.
Cavanna, F. Vissani and L. Coraggio for useful discussions.

\end{document}